\documentclass[a4paper]{jpconf}
\usepackage{amssymb}
\usepackage{graphicx}
\usepackage{amsthm}
\usepackage{latexsym}
\def\half{\textstyle{1\over2}}

\def\R{{\cal{R}}}
\def\E{{\cal{E}}}

\def\r{\mathrm{r}}
\def\z{\mathrm{z}}

\def\R{\mathrm{R}}
\def\Z{\mathrm{Z}}

\newcommand{\bdm}{\begin{displaymath}}
\newcommand{\edm}{\end{displaymath}}

\newcommand{\be}{\begin{equation}}
\newcommand{\ee}{\end{equation}}
\newcommand{\bea}{\begin{eqnarray}}
\newcommand{\eea}{\end{eqnarray}}
\newcommand{\ba}{\begin{eqnarray}}
\newcommand{\ea}{\end{eqnarray}}
\newcommand{\bean}{\begin{eqnarray*}}
\newcommand{\eean}{\end{eqnarray*}}
\newcommand{\bml}{\begin{mathletters}}
\newcommand{\eml}{\end{mathletters}}
\newcommand{\del}{\overrightarrow{\nabla}}

\newcommand{\bb}{\bibitem}

\begin{document}
\title{\textbf{Ernst equation and spheroidal coordinates with a cosmological constant term}}
\author{\textbf{C.Charmousis}}
\date{}
\address{ LPT, Universit\'e de Paris-Sud,\\ B\^at. 210, 91405 Orsay
CEDEX, France\\LPT-0672}
\ead{\textbf{christos.charmousis@th.u-psud.fr}}
 \vspace{1cm}
%\numberwithin{equation}{section}
%%%%%%%%%%%%%%%%%%%%%%%%%%%%%%%%%%%%%%%%%%%%%%%%%%%%%%%
\begin{abstract}
We discuss solution generating techniques treating  stationary and
axially symmetric metrics in the presence of a cosmological
constant. Using the recently found extended form of Ernst's complex
equation, which takes into account the cosmological constant term,
we propose an extension of spheroidal coordinates adapted to
asymptotically de-Sitter and anti de-Sitter static spacetimes. In
the absence of a cosmological constant we show in addition that any
higher dimensional metric parametrised by a single angular momentum
can be given by a 4 dimensional solution and Weyl potentials
parametrising the extra Killing directions. We explicitly show how a
stationary, and a static axially symmetric spacetime solution in 4
dimensions, can be {\it added} together to give a 5 dimensional
stationary and axisymmetric solution.
 \end{abstract}
%%%%%%%%%%%%%%%%%%%%%%%%%%%%%%%%%%%%%%%%%%%%%%%%%%%%%%%

\section{Introduction}

In recent years there has been an increasing effort in finding exact
solutions of higher dimensional gravity \cite{mp}, \cite{ER},
\cite{Harmark}, \cite{ER02}. In particular, since the pioneering
work of Maldacena \cite{maldacena} bringing into perspective the
adS/CFT correspondence some effort has been devoted to understanding
the effect of the cosmological constant term in Einstein gravity
(see for example \cite{hawking}, \cite{Charmousis:2003wm},  \cite{clsz}). Such studies
have also been motivated recently from braneworld
gravity. Using the adS/CFT correspondence in the context of
braneworlds \cite{kaloper},  intriguing relations  between bulk
higher dimensional black holes and their 4-dimensional quantum
versions (see also \cite{frw}) have been put forward. If this correspondance were true, exact
higher dimensional solutions solutions would be providing intriguing
information about the quantum description of 4 dimensional black holes from higher dimensional classical solutions.
However (and not-surprisingly) such solutions have been proven very difficult to find analytically, and in
particular in the presence of a cosmological constant term, which is
vital for an adS/CFT description. It seems that better understanding of solution generating techniques as well as the investigation of convenient coordinate systems, involving the cosmological constant term,
are needed in order to tackle such problems were, even in the case of 4 dimensional general relativity, very little is known.

In a recent paper \cite{clsz}, rotating spacetimes of axial symmetry
were studied in the presence of a cosmological constant. Classical
techniques, as that of Lewis-Papapetrou \cite{lewis}, were
developed there to include the cosmological constant term. In this letter,
we will firstly focus on the Ernst equation \cite{Ernst}, as well as extend on a 
solution generating method developed in \cite{clsz}. The Ernst equation,
extended for a cosmological constant term \cite{clsz},
will permit us here to propose an extension of spheroidal
coordinates to adS/dS static black holes. These coordinates, for
$\Lambda=0$, have been shown to be very useful in the study of
stationary metrics. In particular Ernst \cite{Ernst} showed how one
could generate rather simply Kerr's solution starting from
Schwarzschild. The extension of asymptotically flat coordinate
systems to asymptotically $\Lambda\neq 0$ coordinates maybe very
important in order to find novel stationary solutions such as the
$adS$ version of the black ring solution \cite{ER}.
The latter solution generating method on the other hand will allow
us, using Weyl's classical GR formalism, to extend 4 dimensional
solutions  to higher dimensional ones. In particular we will show
how by literally adding together the potentials of 4 dimensional static and stationary
metrics we can construct 5 dimensional stationary ones.

In 4 dimensional general relativity Einstein's equations in the
vacuum $R_{AB}=0$, guarantee that any locally static and
axially-symmetric metric can be written as
\be
ds^2=-e^{2\lambda}dt^2+e^{-2\lambda}\left[\alpha^2d\varphi^2+e^{2\chi}(d\r^2+d\z^2)\right],
\label{WC} \ee
where the metric components depend on $\r$ and $\z$.
Since the field equation for $\alpha$ reads,
\be
\Delta \alpha=0
\ee
by a suitable 2 dimensional conformal coordinate transformation we can set
$\alpha=\R$ thus obtaining the Weyl form \cite{Weyl}
\be ds^2=-e^{2\lambda}dt^2+e^{-2\lambda}\left[\R^2d\varphi^2
+e^{2\chi}(d\R^2+d\Z^2)\right], \label{WC1} \ee
where $\lambda$, the Weyl potential, and $\chi$ depend on $\R$, $\Z$
and satisfy the field equations,
\bea \label{weylrod} \left(
\partial_{\R}^2 +\frac{1}{\R}\partial_{\R} +\partial_{\Z}^2\right) \lambda=0.\\
\label{nonlinear}
\partial_{\R}\chi=\R\left[\left(\partial_{\R}\lambda\right)^2-
\left(\partial_{\Z}\lambda\right)^2\right], \qquad
\partial_{\Z}\chi=2\, \R\, \partial_{\R}\lambda\partial_{\Z}\lambda
\eea
Equation (\ref{weylrod}) is the linear Laplace equation written in
three dimensional cylindrical coordinates. Given that
(\ref{weylrod}) is a linear equation, a general solution can be found
by simple separation of variables and imposing adequate asymptotic
boundary conditions \cite{quevedo}. In a more pictorial manner, one
can view Weyl potentials $\lambda$, as Newtonian sources in three
dimensions and can, according to (\ref{weylrod}) even superpose them creating new solutions from
known ones. Once the Weyl potential $\lambda$ has been specified,
one evaluates the $\chi$ field by direct integration from
(\ref{nonlinear}). Equations (\ref{nonlinear}) actually carry the
full non-linearity of Einstein's equations.  Weyl potentials are not
unique and are associated to the patch of coordinates we are using.
Flat space for example has Weyl potentials (modulo a constant) given
by,
\be \label{flat} \lambda=0, \quad \lambda= \ln \R, \quad \lambda=
\half \ln (\sqrt{\R^2+\Z^2}+\Z) \ee
where in particular the last one is adapted to an accelerating
Rindler patch. A Schwarzschild black hole of mass $M$ has Weyl
potential given by
\be \label{marion} \lambda=\half \ln \left(\frac{\R_+ + \R_-
-2M}{\R_++ \R_-+2M}\right) \ee
where $\R_\pm= \R^2+(\Z \pm 2M)^2$. As we mentioned one can superpose
black hole Weyl potentials obtaining multiple black hole solutions
\cite{israelkahn}. Typically when sources are superposed, conical
singularities for $\R\rightarrow 0$ appear, and are interpreted as
struts holding, or strings pulling the sources apart thus keeping
the attractive sources in a {\it static} equilibrium.

In essence, Weyl components adapt the problem of finding solutions
to a three dimensional flat coordinate system. This system is
convenient for analysing the solutions in parallel with their
Newtonian sources but is not always tailored to the solutions
themselves. Quite often it is more suitable to adapt the coordinate
system to the Weyl potential $\lambda$ rather than $\alpha$. This is
one of the ideas behind spheroidals although they were not initially
introduced or defined this way. These coordinates  were first
discussed in the context of axial symmetric spacetimes by
Zipoy{\footnote{According to \cite{Zipoy} such coordinates were used
to describe the exact Newtonian gravitational field of the Earth.}}
\cite{Zipoy} (see also \cite{Harmark} for $D\geq 4$). Consider
polar-like coordinates $(u,\psi)$ but with hyperbolae as radial
functions, that is
\bea
\Z&=& \cosh u\, \cos\psi\, , \nonumber\\
\R&=& \sinh u \, \sin\psi\, , \label{prolate} \eea
so that in the $(\R,\Z)$ plane $\psi={\rm const}$ curves are
hyperboloids and $u={\rm const}$ are ellipsoids. On setting $x=\cosh
u$ and $y=\cos\psi$, the coordinate system becomes anew symmetric in
$x$ and $y$.

Consider a Schwarzschild black hole: the standard metric,
%cc
\be \label{julien} ds^2=-\left(1-{2M \over
r}\right)dt^2+\frac{dr^2}{1-{2M\over r}}+r^2 d\Omega_{II}^2 \ee
%cc
can be rewritten in Weyl coordinates $(\r,\z)$ of (\ref{WC}) setting
$r^2/2M=\cosh^2(\r/2)$ and $\theta=\z$. The conformal transformation
to (\ref{WC1}) and (\ref{prolate}) gives \be \label{5}
e^{2\lambda}=\frac{x-1}{x+1} \ee
so that in spheroidals the Weyl potential for (\ref{julien}) is
rather simple. We will come back to this point in a moment but first
let us switch-on rotation.

Lewis and Papapetrou  \cite{lewis} generalised the  approach of Weyl
to stationary and axisymmetric solutions in vacuum. After a
conformal transformation, the metric takes the Lewis-Papapetrou form
\be ds^2=-e^{2\lambda}\left(dt+Ad\varphi\right)^2+e^{-2\lambda}
\left[\R^2d\varphi^2+e^{2\chi}(d\R^2+d\Z^2)\right], \label{stat} \ee
which differs from the static form by  the additional component
$A=A(\R,\Z)$. Note that $\partial_t$ is no longer a static but
rather a stationary locally timelike Killing vector field. For the
metric (\ref{stat}), Ernst \cite{Ernst} pointed out an interesting
reformulation of Einstein's equations for $A$ and $\lambda$, which
read respectively \be \label{ZZ}
\partial_{\R}\left(\frac{e^{4\lambda}}{\R}\partial_{\R}A\right)+\partial_{\Z}\left(\frac{e^{4\lambda}}{\R}\partial_{\Z }A\right)=0,
\qquad \left(
\partial_{\R}^2+\frac{1}{\R}
\partial_{\R}+\partial_{\Z}^2\right) \lambda=\frac{e^{4\lambda}}{2\, \R^2}\left[(\partial_{\R}A)^2+(\partial_{\Z}A)^2
\right]. \nonumber \ee Indeed introduce an auxiliary field,
$\omega$, defined by
\be \label{zoo} \left(-\partial_{\Z}\omega,
\partial_{\R}\omega\right)=\frac{e^{4\lambda}}{\R}
\left(\partial_{\R} A,\partial_{\Z} A\right), \ee
 and the complex function \be \E=e^{2\lambda}+i\omega \label{TPs} \ee
then satisfies the differential equation \be \label{robin}
\frac{1}{\R} \del \cdot \left (\R \del \E \right ) = \frac{(\del
  \E)^2}{\mbox{Re}(\E)}
\label{E4} \ee where $\del=(\partial_\R,
\partial_\Z)$. This complex partial differential equation is known as the Ernst equation \cite{Ernst}. Its real
and imaginary part are exactly (\ref{ZZ}). In this language, the
Weyl potential $\lambda$ is simply given by the real part of the
Ernst potential $\E$, whereas rotation is embodied by a non-trivial
$\omega$ (or $A$).

Using the symmetries of complex functions, several methods have been
proposed to obtain solutions of the Ernst equation (\ref{E4}) and
hence to generate new solutions (see \cite{SKMHH}, \cite{Ernst},
\cite{piotr} and references within). An elegant example appeared in
Ernst's original paper \cite{Ernst}, namely a simple method to
obtain the Kerr solution from the Schwarzschild solution using
spheroidal coordinates. Indeed, consider the Mobius transform,
\be \E=\frac{\xi-1}{\xi+1}. \label{xidef} \ee
defining a new potential $\xi$ which now solves the Ernst equation, 
\be \label{cplxernst}
\frac{1}{\alpha} \del \cdot \left ( \alpha \del \xi \right ) =
    \frac{2\xi^* \left ( \del \xi \right ) ^2}{|\xi|^2 -1} \, ,
\ee where a star denotes complex conjugation.{\footnote{In this
representation of the potential, (\ref{cplxernst}) is invariant
under the complex transformation $\xi \rightarrow \xi
e^{i\vartheta}$ for any phase $\vartheta \in R$ and stationary
solutions can be easily generated from static ones (see \cite{Ernst}
and \cite{clsz} for more details).}} It follows immediately from
(\ref{5}) and (\ref{TPs}) that $\xi=x$ for (\ref{julien}). Hence we
notice that our transformed Ernst potential $\xi$ is now the new
'radial' coordinate $x$. This is in contrast to (\ref{WC}) where
$\alpha=\R$ and the Ernst potential is given by (\ref{marion}). In
other words we have adapted the coordinate system to the real part
of the black hole Ernst potential. We will be using this as our
starting definition for extending spheroidal coordinates when we
will switch on the cosmological constant term. Just in order to
complete the discussion on the construction of Kerr, given the
$x\leftrightarrow y$ symmetry, $\xi=y$ is also solution of
(\ref{E4}), as is $\xi=x \sin\vartheta +i y \cos\vartheta$. It turns
out that this is nothing other than the Ernst potential of the Kerr
black hole, where $\sin\vartheta= a/M$ is the ratio between the
angular momentum parameter and the mass of the black hole
\cite{Ernst}.

When we consider a $D$-dimensional spacetime with a cosmological
constant,  a rotating metric of axial symmetry and with a single
component angular momentum can be conveniently written in the form
\cite{clsz},
\ba \label{melina} ds^2 &=& e^{2\nu}
\alpha^{-\frac{D-3}{D-2}}(dr^2+dz^2)+\alpha^{2\over
  {D-2}}
\left[e^{-\sqrt{\frac{D-4}{2(D-2)}}\Psi_0}\left[ -e^{\Omega\over
2}(dt+A
  d\varphi)^2+e^{-\Omega\over 2}
  d\varphi^2\right]\right.+ \nonumber \\
 &&+ \left.e^{\sqrt{\frac{2}{(D-2)(D-4)}}\Psi_0}
  \sum_{i=1}^{D-4}e^{2\Psi_i}(dx^i)^2\right] \, .
\ea
We have $(D-2)$ Killing vectors but as for $D=4$, $\Lambda=0$ only
two of them are not orthogonal, $\partial_t$ and $\partial_\phi$.  The
fields $\alpha$, $\nu$, $\Omega$ and $\Psi_\mu$ again depend on $r$
and $z$. These metric components extend the Lewis-Papapetrou form of
the previous section \cite{clsz}. Indeed the field equations take
the form
\ba \label{PD1}
\Delta \alpha &=& -2 \Lambda \alpha^{\frac{1}{D-2}} e^{2\nu} \\
\label{PD2}
0 &=& \overrightarrow{\nabla} \cdot \left ( e^{\Omega}
  \alpha \overrightarrow{\nabla} {A} \right )  \\
\label{PD3} \frac{1}{\alpha} \overrightarrow{\nabla} \cdot \left (
\alpha \overrightarrow{\nabla} {\Omega} \right ) &=& 2 \,
\epsilon \, e^{\Omega} \left (\overrightarrow{\nabla} A \right )^2 \\
\label{PD4}
\overrightarrow{\nabla} \cdot \left ( \alpha
  \overrightarrow{\nabla} {\Psi_\mu} \right ) &=&  0
,\qquad \mu=0...d-3
\\
\label{PD5}
2 \nu_{,u} \frac{\alpha_u}{\alpha}- \frac{\alpha_{,uu}}{\alpha} &=&
\frac{1}{4}
\left(\Psi_{0,u}^2 +\frac{1}{2}  \Omega_{,u}^2 \right)
+\frac{\epsilon}{2} e^{\Omega} \left ( A_{\, ,u} \right )^2
+\sum_{i=1}^{D-4}\Psi_{i,u}^2, \qquad
(u\leftrightarrow v)\\
\ea
If $\Lambda\neq 0$, $\alpha$ is no longer a harmonic function.
Furthermore, $\alpha$ fixes $\nu$ from (\ref{PD1}) which is no
longer directly given by (\ref{PD5}) as for $\Lambda=0$. Pictorially
$\alpha$ is a messenger component relating all the equations
together. This is in contrast to the case when $\Lambda=0$ and
equations (\ref{PD4}) are decoupled from (\ref{PD2}) and (\ref{PD3})
(see the geometric interpretation in \cite{clsz}). The components
$\Psi_\mu$ and $\Omega$ play a similar role to the Weyl potential
$\lambda$ of (\ref{WC1}). When $D=4$ we have by definition
$\Psi_\mu=0$. It will be useful here to rewrite the field equations
in terms of the dual potential $\omega$,
\be \label{coffee} (-\partial_z\omega,
\partial_r\omega)=e^{\Omega} \alpha
(\partial_r A, \partial_z A). \ee
As it was demonstrated in \cite{clsz}, using (\ref{coffee}), we can
rewrite (\ref{PD2}), (\ref{PD3}) in terms of a single complex
differential equation with respect to the complex potential
\be \E=e^{\frac{\Omega}{2}}\alpha+i\omega,
 \label{Em1}
\ee
In fact, we can go one step further and rewrite the field equations
with respect to $\E$. We get
\ba \label{PE1} \mbox{Re}(\E) \Delta \alpha   &=& -2 \Lambda e^{2\bar{\nu}}\alpha^{\frac{D}{2(D-2)}} \\
\label{PE2} \frac{1}{\alpha} \del \cdot \left (\alpha \del \E \right
) &=& \frac{(\del \E)^2}{\mbox{Re}(\E)} + \mbox{Re}(\E)
\frac{\Delta \alpha}{\alpha} \\
\label{PE3} \overrightarrow{\nabla} \cdot \left ( \alpha
  \overrightarrow{\nabla} {\Psi_\mu} \right ) &=&  0
,\qquad \mu=0...D-4
\\
\label{PE4} 2 \bar{\nu}_{,u} \frac{\alpha_u}{\alpha}-
\frac{\alpha_{,uu}}{\alpha} &=& \frac{1}{4} \Psi_{0,u}^2 +
\half\frac{\E_{,u}\E_{,u}^{*}}{{\mbox{Re}(\E)}^2}+
\sum_{i=1}^{D-4}\Psi_{i,u}^2 \quad (u\leftrightarrow v)\\
\ea
where we have redefined for convenience $2\nu=2\bar{\nu}-\half \ln
\alpha -\frac{\Omega}{2}$. Equation (\ref{PE2}) extends the Ernst
equation \cite{Ernst} to the presence of a cosmological
constant-indeed the extra term in $\Delta\alpha$ drops out if
$\Lambda=0$ and we get (\ref{robin}).

We stress here that Weyl coordinates (\ref{WC1}) cannot be used once
$\Lambda\neq 0$. We cannot therefore set $\alpha=\R$ and the
equations (\ref{PE1}-\ref{PE4}) are no longer integrable.  We
noticed however,  that if we consider (\ref{xidef})
 then the Ernst potential of the Schwarzschild black hole
(\ref{julien}) is simply $\xi=x$ in spheroidal coordinates. Given
that we  have the extension of Ernst's equation for $\Lambda\neq 0$
we can now consider doing the same trick for $\Lambda\neq 0$.
Consider first the four dimensional Kottler black hole with line
element
\be \label{schw} ds^2=r^2\left({dr^2\over r^2
V(r)}+d\theta^2\right)- V(r) dt^2 +r^2 \sin^2 \theta d\psi^2 \ee and
$V(r)=1+k^2r^2-{2M \over r}.${\footnote{In the presence of a
negative cosmological constant the generalised metric reads,
$ds^2=-Vdt^2+{dr^2\over V(r)} +r^2 dK^2_{D-2}$ with $V(r)=\kappa+k^2
r^2 - {2M\over r^{D-3}}$, adS curvature scale
$2\Lambda=-(D-1)(D-2)k^2$ and constant curvature of the $D-2$
compact space $\kappa=0,1,-1$. Note that in  the presence of a
negative cosmological constant we can obtain a black hole geometry
with a flat (compact) horizon-the cosmological constant providing
the necessary curvature scale. Furthermore, switching-off $M$ gives
us different slicings of constant curvature spacetime.}
Note that we will bypass Weyl coordinates by setting $\frac{dr}{r
V(r)}=d\r$ and not explicitly do the integral which involves
elliptic functions. Comparing with (\ref{melina}) we now read off
the metric components in $(r,\theta)$,
\be \alpha =r \sin\theta  \sqrt{V},\quad \E = V, \quad \e^{2\nu} =
r^2 \alpha^{1/2}. \ee
and therefore we set
\be \frac{x-1}{x+1}=V(r)=>x=-1-\frac{2}{k^2r^2-{2M\over r}}
\label{nathan}\ee
On the other hand setting as before $y=\cos\theta$ the metric
(\ref{schw}) takes the form
\be
ds^2=-\frac{x-1}{x+1}dt^2+r^2\left[\frac{dx^2}{(x^2-1)(x+1)^2}\frac{1}{(r^2
k^2+{2M\over r})^2} +\frac{dy^2}{1-y^2}+(1-y^2)d\phi^2\right] \ee
and $r$ is the real positive root of the third order polynomial,
$k^2r^3+\frac{2}{x+1}r-2M=0$ with respect to the new radial
coordinate $x$. Note from (\ref{nathan}) that there is an additional coordinate singularity when $V=1$. We will come back to this for the more feasible case of $D=5$. If we set $k=0$ we recover the usual spheroidal
patch of (\ref{schw}) \cite{Harmark}. It is of more interest here to
set $M=0$ in order to study adS and dS. For the former case we start
with the global patch of adS and setting $X=-x$ we get,
\be 
\label{isa}
ds^2=-\frac{X+1}{X-1}dt^2 + \frac{dX^2}{2k^2
(X+1)(X-1)^2}+\frac{1}{X-1} \left[
\frac{dy^2}{1-y^2}+(1-y^2)d\phi^2\right] \ee
The coordinate range is $X>1$, $-1<y<1$ and only covers half of anti
de-Sitter space with the boundary sitting at $X=1$. The coordinate
transformation for planar or hyperbolic slicings of adS goes through
the same way. For $\Lambda=0$ we saw that in the $(\R,\Z)$-plane the
integral curves $x=constant$ are hyperbolas. So what happens for
adS? To see this set $kr= \cosh{ku}$ which takes us to the usual
global adS patch. Note now from the form of (\ref{WC}) that the Weyl
coordinate $\r$ reads,
\be \r=\ln |\tanh\frac{ku}{2}| \ee
which means that after a conformal transformation according to
(\ref{prolate}) we have $X=\coth(ku)$ and $y=cos(\theta)$. The
integral curves are therefore similar but as we noticed earlier we
need two patches to cover $u\in ]-\infty,+\infty[$ with a branch cut
at $u=0$.  The adS boundary is then at $X=1$ and at $X=-1$ for the
second patch. For de-Sitter set $k^2=-a^2$ starting from the locally
static patch (\ref{schw}) to get:
\be ds^2=-\frac{x-1}{x+1}dt^2 + \frac{dx^2}{2a^2
(x+1)(x^2-1)}+\frac{1}{1+x}
\left(\frac{dy^2}{1-y^2}+(1-y^2)d\phi^2\right)
 \ee
The coordinate range is $x>-1$ and the usual horizon at $r=1/a$ is
not in this range. Therefore the spheroidals cover a yet smaller
patch than the usual static de-Sitter (\ref{julien}).

For $D=5$ there is an interesting twist. Proceeding as before we get
\be \E=V(r) (r\cos\theta) \ee
and now $\E$ is a function of $r$ and $\theta$. Note however that
$\E_0=r\cos\theta$ is simply the Ernst potential for flat spacetime
(i.e. when we set $k=0$ and $M=0$) and we can therefore neglect it
keeping the Ernst potential modulo flat spacetime (this also agrees
with  \cite{MI} for $D=5$ and $\Lambda=0$). In other words we once
more have (\ref{nathan}) (for $D=5$) which yields, \be
x+1=-\frac{2}{k^2r^2-{2M\over r^2}}. \ee Note that there is an extra coordinate singularity at $r_S^4={\mu\over k^2}$ (or $V=1$) which is to the right of the event horizon at $r=r_H$ or $x=1$. This singularity goes to asymptotic infinity as the cosmological constant goes to zero $k\rightarrow 0$ and dissapears for the flat case. Close inspection therefore reveals that for $r>r_H$ there are two branches in spheroidal coordinates, one for $r_H<r<r_{S^{-}}$, $x>1$ and one for $r>r_{S^{+}}$, $x<-1$. Setting 
$y=\cos2\theta$ the first branch reads
\bea
ds^2=-\frac{x-1}{x+1}dt^2+\frac{\sqrt{1+2Mk^2(x+1)^2}-1}{k^2(x+1)}
\left[\frac{dx^2}{4(x^2-1)(1+2Mk^2(x+1)^2)}+ \right.\nonumber \\
\left. +\frac{dy^2}{4(1-y^2)}+(1-y)d\phi^2+(y+1)d\psi^2 \right] \eea
and describes the region close to the black hole event horizon. In particular the limit $M\rightarrow 0$ is ill-defined in this branch whereas we have a smooth $k\rightarrow 0$ limit. For the second branch setting as before $X=-x$ (ie. $X>1$) and using the same angular coordinate,
\bea
ds^2=-\frac{X+1}{X-1}dt^2+\frac{\sqrt{1+2Mk^2(X-1)^2}+1}{k^2(X-1)}
\left[\frac{dX^2}{4(X^2-1)(1+2Mk^2(X-1)^2)}+ \right.\nonumber \\
\left. +\frac{dy^2}{4(1-y^2)}+(1-y)d\phi^2+(y+1)d\psi^2 \right] \eea
This branch has the correct limit of $M\rightarrow 0$ giving the equivalent of (\ref{isa}) in $D=5$ and describes the region close to the adS boundary.
An interesting open question, regarding the extension of spheroidals
we have undertaken here, is if they can give the rotating
generalisation of (\ref{julien}) as so happens for $\Lambda=0$
\cite{Ernst}, \cite{MI}.

As observed in \cite{clsz} all equations
(\ref{PE2}-\ref{PE4}) apart from (\ref{PE1}) are independent of $D$
whereas of course the metric (\ref{melina}) depends on the
dimension. Therefore starting from a $D=4$ stationary and axisymmetric 
solution with $\Lambda=0$
we can construct an infinite number of $(D+n)$-dimensional solutions
parametrised by the $n$ extra Weyl potentials $\Psi_\mu$-solutions
of (\ref{PE3}). In fact we have
\be \label{basic} 2 (\nu_{(D),u}-
\nu_{(D+1),u})\frac{\alpha_u}{\alpha}= -\frac{1}{4} \Psi_{u}^2,
\qquad (u\leftrightarrow v) \, . \ee
where the subscript $D$ refers to the spacetime dimension for $\nu$.
Setting $\sigma=\nu_{(D),u}- \nu_{(D+1),u}$ we rewrite the above
equation in terms of $\R$ and $\Z$ in Weyl coordinates:
\bea 4 \sigma_{,\Z}=-2 \R \Psi_{,\R}\Psi_{,\Z} \nonumber\\
8\sigma_{,\R}=\R (\Psi_{,\R}^2-\Psi_{,\Z}^2) \label{sphere}\eea
In \cite{clsz} examples were given where one started from a higher
dimensional solution and found its seed $D=4$ solution. Here we will
do the opposite and uplift solutions making use of the Weyl
formalism \cite{Weyl}. Consider as our seed metric in $D=4$ that of
Kerr,
\bea \label{carter}
ds_4^2&=&-\frac{\Delta}{\rho^2}\left(dt-a\sin^2\theta
d\varphi\right)^2
+\frac{\sin^2\theta}{\rho^2}\left(a dt-(r^2+a^2)d\varphi\right)^2\nonumber\\
&+& \rho^2 \left({dr^2\over \Delta
  }+d\theta^2\right),
\eea
where $M$ is the black hole mass, $a$ the angular momentum parameter
and \be \Delta = r^2+a^2-2M r, \quad \rho^2 = r^2+a^2\cos^2\theta.
\ee Comparing with (\ref{melina}) we now read off the relevant
components
\bea \alpha &=& \sin\theta, \quad \e^{2\nu}= \rho^2 \alpha^{1/2}
\sqrt{\Delta},\\
A&=& {2Ma\sin^2 \theta \over {\rho^2-2M}},\quad
e^{\Omega}=\frac{ (\rho^2-2M)^2}{\Delta \rho^4 \sin^2 \theta }. \\
 \eea
In order to uplift the solutions we will use  Weyl coordinates
(\ref{WC1}) which are related to (\ref{carter}) by the relations
\be \label{isabelle} \R=\sqrt{\Delta} \sin\theta =\alpha, \quad
\Z=(r-M)\cos\theta. \ee
Note that if we set $\lambda=\Psi/\sqrt{6}$ in $\ref{PE3}$ and $\chi=-4\sigma /3 $
in (\ref{sphere}) we obtain the 4 dimensional Weyl equations,
(\ref{weylrod}) and (\ref{nonlinear}) which are solutions of the static and axially symmetric metric
(\ref{WC1}). We have thus demonstrated the following: for each seed
solution of $\Lambda=0$, (\ref{PE1}-\ref{PE4}), for our example here
(\ref{carter}), we can take any Weyl solution in 4 dimensions (in
the form (\ref{WC1})), use the coordinate transform relating it to
the seed coordinate system (here (\ref{isabelle})) and thus obtain a
$D=5$ solution given by,
\be\label{sol} ds^2=\rho^2 \R^{-\frac{1}{6}}
e^{\frac{3\chi(\R)}{2}}(dr^2+dz^2)+\R^{2/3}
e^{-\lambda(\R)}\left[\left[ -e^{\Omega\over 2}(dt+A
  d\varphi)^2+e^{-\Omega\over 2}
  d\varphi^2\right]+ \R^{2/3}e^{2\lambda(\R)} d\psi^2\right]\ee
We emphasize that $\R$ and $\Z$ are functions of $r$ and $z$ ie
verify (\ref{isabelle}).

Lets consider a flat potential (\ref{flat}) as an example. Start by
noting that (\ref{weylrod}) has the obvious symmetry,
$\lambda\rightarrow l \lambda,\mbox{ }l\in R$. As it was pointed out
in \cite{clsz}, generically, when uplifting solutions asymptotic
flatness is not guaranteed. Indeed in 5 dimensions, the extra
dimensional $d\psi^2$ component in (\ref{sol}) reads $\R^{2/3}
e^{2\lambda}$ and given the form of $\R$ in (\ref{sphere}) an
obvious problem will be to get this extra direction asymptotically
flat. So lets start with $\lambda=l \ln \R$ where $l\in \R$. It is
straight forward to integrate and we get
\bea ds^2=\rho^2 \R^{\frac{9 l^2-1}{6}}(dr^2+dz^2)+
\R^{\frac{2}{3}-l}\left[\left[ -e^{\Omega\over 2}(dt+A
  d\varphi)^2+e^{-\Omega\over 2}
  d\varphi^2\right]+ \R^{\frac{2}{3}+2l} d\psi^2\right] \eea
This is a solution to Einstein's equations for all $l$ where one
uses (\ref{isabelle}).
 For example, choosing
$l=-1/3$ we get $\R^{2/3}e^{2\lambda}=1$. It is then easy to check
that this is the Kerr string solution in 5 dimensions. One can
follow the same technique using this time the Weyl potential for the
static black hole, (\ref{marion}) and adjusting the free constants in
such a way as to obtain an asymptotically flat spacetime. It would
be particularly interesting to extend this method to the case where
$\Lambda\neq 0$. By analogy to what we have done here, it is
possible that a starting point for such a generalisation would be to
find how the additional fields of (\ref{PE1}-\ref{PE4}) transform
upon keeping the Ernst potential $\E$ fixed.

\ack It is a great pleasure to thank my collaborators in \cite{clsz}
David Langlois, Dani\`ele Steer and Dr. Robin Zegers as well as the
organisers of the XBII conference. I also thank Roberto Emparan and Ruth Gregory 
for discussions on spheroidals. I finally thank the Galileo
Galilei Institute for Theoretical Physics for their hospitality and
the INFN for partial financial support.

\end{document}